\documentclass[11pt,aps,notitlepage,onecolumn,superscriptaddress,showpacs,nofootinbib,longbibliography,amssymb]{revtex4-1}

 \usepackage{amsmath}
\usepackage{amsfonts}
\usepackage{amssymb}
\usepackage{amscd}
\usepackage{accents}
\usepackage{marvosym}
\allowdisplaybreaks[4]

\usepackage{bm}
\usepackage{titlesec}
\titleformat{\section}{\centering\large\sffamily\bfseries}{\thesection.}{1em}{}
\titleformat{\subsection}{\centering\large\sffamily}{\thesubsection.}{0.3em}{}
\titleformat{\subsubsection}{\centering\sffamily}{\thesubsection.\thesubsubsection.}{0.3em}{}

 \usepackage[colorlinks]{hyperref}
 \usepackage[capitalise]{cleveref}
\usepackage{xcolor}
\definecolor{dark-red}{rgb}{0.6,0.15,0.15}
\definecolor{dark-blue}{rgb}{0.15,0.15,0.8}
\definecolor{medium-blue}{rgb}{0,0,0.6}
\hypersetup{
    colorlinks, linkcolor={dark-red},
    citecolor={dark-blue}, urlcolor={medium-blue}
}

\begin{document}
\title{An approximate global solution of Einstein's equation for a rotating compact source with linear equation of state}

\author{J. E. Cuch\'i}
  \affiliation{Dpto. F\'isica Fundamental, Universidad de Salamanca}
  \thanks{J. E. Cuch\'i(\Letter): \texttt{jecuchi@usal.es}
E. Ruiz: \texttt{eruiz@usal.es} \\
A. Gil-Rivero: \texttt{tonipungun@hotmail.com}}
 \author{A. Gil-Rivero}
 \affiliation{Dpto. F\'isica Fundamental, Universidad de Salamanca}
 \author{A.\ Molina}
 \affiliation{Dpto. F\'isica Fonamental,  Institut de Ci\`encies del Cosmos, Universitat de Barcelona}
 \thanks{A. Molina: \texttt{alfred.molina@ub.edu}}
 \author{E.\ Ruiz}
 \affiliation{Dpto. F\'isica Fundamental, Universidad de Salamanca}

\begin{abstract}
 We use analytic perturbation theory to present a new approximate metric for a rigidly rotating perfect fluid source with equation of state (EOS) $\epsilon+(1-n)p=\epsilon_0$. This EOS includes the interesting cases of strange matter, constant density and the fluid of the Wahlquist metric. It is fully matched to its approximate asymptotically flat exterior using Lichnerowicz junction conditions and it is shown to be a totally general matching using Darmois-Israel conditions and properties of the harmonic coordinates. Then we analyse the Petrov type of the interior metric and show first that, in accordance with previous results, in the case corresponding to Wahlquist's metric it can not be matched to the asymptotically flat exterior. Next, that this kind of interior can only be of Petrov types I, D or  (in the static case) O and also that the non-static constant density case can only be of type I. Finally, we check that it can not be a source of Kerr's metric.
\end{abstract}

\maketitle

\section{Introduction}
One of the regrettable facts about General Relativity is that, by now, despite the numerous exact solutions and modern methods to generate them, it has not been able to provide an exact solution describing a rotating stellar model, i.\,e.\!\! a spacetime corresponding to an isolated self-gravitating rotating fluid in equilibrium, other than the rotating  disc of dust described by Neugebauer and Meinel in \citep{neugebauer1995general} and its generalization for counter-rotating discs \citep{klein2001exact}. Although this infinitesimally thin disc solutions are useful models for galaxies and accretion discs, they are quite far from describing sources of spheroidal shape which are the most common astrophysical objects.

Stellar models are built matching an interior spacetime describing the source and the exterior spacetime that encloses it. A candidate interior solution should correspond to a stationary axisymmetric perfect fluid without extra symmetries and admit a zero pressure surface. While the Einstein equations for stationary and axisymmetric exteriors form a completely integrable system \citep{maison1978are,maison1979complete} and then can be dealt with using solution generation methods to get general solutions, interiors are far more complicated. The only case we know to form a completely integrable system is the disc of dust so in any other case one can only try to get  particular solutions. In spite of the effort and interest put in the problem, it has proved difficult to obtain non-singular solutions of this kind. To our knowledge, the only candidates have been for a long time the Wahlquist metric \citep{wahlquist1968isf,wahlquist1992problem} and the differentially rotating solution by Chinea and Gonz\'alez-
Romero \citep{
chinea1990interior}. The zero-pressure surface of the latter has finite area but can not enclose the symmetry axis, and while numerical relativity predicts stationary toroidal sources \citep{ansorg2009solution} that can be obtained starting from a spheroidal topology for a sufficiently strong degree of differential rotation, in the case of rigid rotation they are unreachable \citep{ansorg2004equilibrium}. Here we will put our focus on rigidly rotating and --since we are interested also in the static limit of our stellar models-- spheroidal sources.

 Summing to the difficulty of finding suitable interiors, there are the ones arising from the matching with the asymptotically flat exterior. For stellar models it is an overdetermined problem \citep{mars1998cgm} so in general we can not find an exterior that matches a given interior. Such seems to be the case for Wahlquist, where the derivations of the impossibility of matching it with an asymptotically flat exterior \citep{wahlquist1968isf,bradley2000wmc,sarnobat2006wes} come from the analysis of the shape of its surface and involve approximations. This situation leads, if one is to extract information about spheroidal stellar models, to approximate solutions.

Within the field of approximations, one has to choose between the accuracy  and closeness to the real physical problem that numerical methods now provide (in fact, some numerical codes like AKM \cite{ansorg2002highly,ansorg2003highly} are actually ``exact'' in the numeric sense, i.\,e. they reach machine accuracy) and the density of information and greater flexibility for theoretical work that analytic perturbation theory offers. Both fields have successfully treated several important problems along the years in the context of stellar models and binary systems. Among them, the still fresh revolution in the field of coalescence of black holes or compact stars, a problem of fundamental importance for the modern experiments of gravitational waves detection, is actually a collaborative effort of analytic and numerical groups \citep{lrr-2003-3,lrr-2006-4,lrr-2007-2,lrr-2011-6}. A recent application of numerical and analytic approximations to stellar models can be found in \citep{bradley2009quad}.

In \citep{cabezas2007ags} (CMMR from now on) some of us presented a new analytic approximation scheme focused on the obtention of stellar models. It is mainly a post-Minkowskian treatment with a secondary ``slow rotation'', or --to avoid confusion with the conventional use of the name-- more precisely a small deformation approximation. 
 The latter is in general very little restrictive since for the typical densities of compact stars very high ($\sim 0.1$\,kHz) rotation rates are needed to deform sources between one and two solar masses \citep{nozawa1998construction,cuchi2013comparingRevTeX}, the mass most of these objects seem to have \citep{lattimer2011neutron}.
CMMR is suited to static or rigidly rotating fluids with a barotropic equation of state (EOS) and provides the fully matched exterior and interior spacetimes in harmonic gauge. This is actually the main strength of the scheme. There are methods to obtain analytic approximate exterior metrics (see, e.\,g. \citep{teichmuller2011analytical}), but the matching with an interior gives two main advantages besides the knowledge of the interior itself. First, a given stellar model interior can only be matched with at most one exterior \citep{mars1998cgm}. This  drastic reduction of freedom allows that in CMMR, once the EOS is fixed, the whole global spacetime is characterised by only two parameters which can be chosen or obtained from its physical properties. Second, the matching fully determines the zero-pressure surface of the source. Some of the main relevant properties of the exterior from the observational point of view, such as equatorial and polar redshifts and the existence or height of an innermost stable 
circular orbit, depend on the surface and thus increase the interest of global models.
Finally, in general, the CMMR scheme is quite systematic and can be easily iterated to give better precision.

In this paper we apply CMMR to obtain the global solution for a non-convective fluid with EOS $\epsilon+(1-n)p=\epsilon_0$, where $\epsilon,\, p$,  $\epsilon_0$ and $n$ are energy density, pressure, a constant with units of energy density and a free parameter, respectively. Among the relevant EOS included in this family are the ones corresponding to constant density ($n=1$), the Wahlquist and Whittaker fluid ($n=-2$) and $n=4$ which corresponds to the simple MIT bag model EOS that has been frequently used  to study the properties of strange quark matter.  The latter as constituent of at least a class of compact stars is currently an exciting possibility in astrophysics \citep{weber2005strange,weissenborn2011quark,lattimer2011neutron}. We also show that the full matching procedure used in CMMR 
 can be made completely general using Darmois matching conditions \citep{darmois1966memorial} instead of Lichnerowicz ones \citep{lichnerowicz1955theories}, as well as ensuring the 
generality of the 
assumptions made on the embeddings of the matching surface. Later, we use the approximate metric to exclude as candidate sources of stellar models those Petrov type II interiors with this EOS that admit a CMMR expansion. This is unexpected. This fact coincides 
with a lack of exact stationary axisymmetric perfect fluid solutions of that Petrov type \cite{senovilla1993contribucion}, what leads us to wonder if this can be the case irrespective of the EOS. The $\bm{Q}$ matrix of a metric 
with Papapetrou's structure like ours rules out Petrov type III, and the combination of its symmetries and EOS discard type N as well \citep{carminati1988type} and hence we get that the possible Petrov types of matchable rotating interiors 
with this EOS are I or D. Our analysis, though approximate, shows as well that the stationary (non-static) constant density case can only have Petrov type I. Finally, we obtain the conditions for our interior to correspond to an approximate Wahlquist metric and recover the result that Wahlquist can not be matched with stationary axisymmetric asymptotically flat exteriors \cite{bradley2000wmc,sarnobat2006wes}. We also show that our interior can not be a source of Kerr metric.

\section{Interior Solution}
Our objective is to build an approximate stellar model within the context of General Relativity, i.\,e., a global spacetime $({\cal {M}},  \bm{g})$, where $\mathcal{M}$ is a differentiable manifold and $\bm{g}$ a Lorentzian metric, corresponding to a stationary and axisymmetric spacetime containing a compact and rotating perfect fluid surrounded by asymptotically flat vacuum. In practice, this is accomplished obtaining the interior spacetime $({\cal V}^-,\bm{g}^-)$ that contains the perfect fluid and matching it through its boundary with the exterior vacuum solution $({\cal V}^+,\bm{g}^+)$. First, we deal with the solution for the perfect fluid interior $({\cal V}^-,\bm{g}^-)$.

\subsection[Source]{Source characterization}\label{source}

In the interior spacetime $({\cal V}^-,\bm{g}^-)$ let us call $\bm {\xi}^-$ the time-like Killing vector field associated to the stationarity. Let also $\bm{\zeta}^-$ be  the space-like Killing vector field  corresponding to axisymmetry, i.e., a Killing  field with closed orbits and vanishing module on the symmetry axis \cite{mars1993axial}. These two vector fields commute \cite{carter1970cps}, 
\begin{equation}
 \left[\bm{\xi}^-,\bm{\zeta}^-\right]=0\,.
\label{eq1}
\end{equation}
and thus two coordinates in $\cal V^-$ can be adapted to the symmetries. Let them be $\{t,\phi\}$, so we have 
\begin{equation}
 \bm{\xi}^-=\partial_t\qquad\text{and}\qquad \bm{\zeta}^-=\partial_\phi.
\label{eq2}
\end{equation}
We assume the fluid flows on the 2-surfaces generated by the Killing vector fields so its 4-velocity is
\begin{equation}
 \bm{u}=\psi\left(\partial_t+\omega \partial_\phi\right)
\label{eq3}
\end{equation}
with $\psi$ the normalization function and $\omega$ its angular velocity as seen by an observer at infinity.
If the fluid has no energy flux this is equivalent to absence of convective motion  \cite{carter1969kho}. It also implies the verification of the circularity condition \cite{papapetrou1966cgs,carter1973black} and then the integrability of 2-planes everywhere orthogonal to the transitivity surfaces of the isometry group, what allows the metric to be block diagonal (\textsl{Papapetrou's structure}) in  a certain set of coordinates  $\{t,\phi, r, \theta\}$ \cite{kundt1966orthogonal}. The function $\omega$ is taken here constant, so the fluid is rigidly rotating and thus free of expansion and shear.

We will study a fluid with equation of state (EOS) $\epsilon+(1-n)p=\epsilon_0$, where $\epsilon$ is energy density, $p$ is pressure, $n$ is a real number and $\epsilon_0$ is a constant with units of energy density. This linear barotropic EOS contains the ones of two significant exact solutions: $n=1$ corresponds to constant density of the Schwarzschild interior and $n=-2$ gives the EOS of the Wahlquist family of metrics \cite{wahlquist1968isf}.

The assumptions of perfect fluid, circularity and rigid rotation allows us to integrate the Euler equation $\nabla_\alpha T_\beta^\alpha=0$ \cite{boyer1965rfm} to give, for $n\neq0$,
\begin{align}
 \epsilon&=\frac{\epsilon_0}{n}\left[(n-1)\left(\frac{\psi}{\psi_\Sigma}\right)^n+1\right]\qquad\text{and}\label{eq5}\\
p&=\frac{\epsilon_0}{n}\left[\left(\frac{\psi}{\psi_\Sigma}\right)^n-1\right]\,,\label{eq6}
\end{align}
where $\psi_\Sigma$ is the value of the normalization factor $\psi$ on the surface of zero pressure.


\subsection{Approximations and solution}


 In the remainder of this section we apply the scheme developed in \cite{cabezas2007ags} to the present EOS, summarizing its main ideas. 

Assuming that the gravitational field is weak enough, it makes sense to decompose the exact metric $g_{\alpha\beta}$ as
\begin{equation}
 g_{\alpha\beta}=\eta_{\alpha\beta}+h_{\alpha\beta}\,.
\label{eq9}
\end{equation}
Here, $\eta_{\alpha\beta}$ is the  Minkowski metric and $h_{\alpha\beta}$, the deviation from flat geometry, collects all the terms of the expansion
\begin{align}
 h_{00}&=\lambda h_{00}^{(1)}+\lambda^2 h_{00}^{(2)}+\cdots\\
 h_{ij}&=\lambda h_{ij}^{(1)}+\lambda^2 h_{ij}^{(2)}+\cdots\\
 h_{0i}&=\omega(\lambda h_{0i}^{(1)}+\lambda^2 h_{0i}^{(2)}+\cdots)
\label{eq10}
\end{align}
where $\lambda$ is a dimensionless parameter related with the strength of the field. We choose it as $ \lambda:=m/r_0$ with $m:=\frac43\pi\epsilon_0 r_0^3\,,$ where $r_0$ is the coordinate radius of the static spherical fluid mass, and in the term $h_{0i}$ we have included a global $\omega$ factor to get a static metric when $\omega=0$. Note that even after fixing the gauge, the rescaling freedom in the radial coordinate will affect the possible values of $r_0$. This does not affect the capacity of CMMR to give explicit values for the metric and other quantities as we will discuss later and has already been done in \citep{cuchi2012compaProcRevTeX,cuchi2013comparingRevTeX}

This post-Minkowskian approximation will eventually give us a solution metric in terms of a tensor spherical harmonics expansion. To have a cut-off for this expansion, we introduce a slow rotation approximation. We need a rotation-related parameter measuring the degree of deformation of the fluid but it must be noted that, if we want $\psi=\psi_{\Sigma}$ to define almost spherical surfaces instead of cylindrical ones, we need $\omega^2\sim\lambda$ at least. These two requirements led some of us to introduce in \citep{cabezas2007ags}
\begin{equation}
\Omega^2:=\frac{r_0 \omega^2}{m/r_0^2},
\label{omegadef}
\end{equation}
the ratio between the classical centrifugal and gravitational energies (also used by e.g. \citep{hartle1967slowly}), as the slow-rotation parameter. This kind of relation is expectable if we want to deal with gravitationally bounded systems. It also causes that $h_{0i}$ terms, naturally possessing odd powers of $\Omega$ to give the expected behaviour under reversal of rotation direction, have an expansion of the form $h_{0i}=\lambda^{3/2} h_{0i}^{(1)}+\lambda^{5/2} h_{0i}^{(2)}
+{\cal{O}}(\lambda^{7/2})$.

We fix the gauge
to be harmonic, and label these coordinates as $\{x^\alpha\}$. Then, to get $h_{\alpha\beta}$ we have to solve the Einstein's equations and the harmonic conditions system. The post-Minkowskian expansion allows us to solve them iteratively. 
The $i$-th iteration gives the $\mathcal{O}(\lambda^i)$ part of the metric deviation, $\bm{h}^{(i)}$ 
by solving the system of PDEs
\begin{equation}
 \left\{
\begin{aligned}
&\triangle h^{(i)}_{\alpha\beta} = -16\pi \mathcal{T}^{(i)}_{\alpha\beta}+2 N^{(i)}_{\alpha\beta} - \partial_\alpha
K^{(i)}_\beta -
\partial_\beta K^{(i)}_\alpha\,, \\
&\partial^k \left[h^{(i)}_{k\alpha} -\frac12 h^{(i)}\,\eta_{k\alpha}\right] = -K^{(i)}_\alpha\;, 
\end{aligned}
\right.
\label{eq11}
\end{equation}
where $\triangle$ is the flat Laplacian, $\mathcal{T}_{\alpha\beta}=T_{\alpha\beta}-\frac12g_{\alpha\beta}T$ with $T_{\alpha\beta}$ the energy-momentum tensor and $N_{\alpha\beta}$ and $K_{\alpha}$ collect quadratic and higher order terms of the Ricci tensor and the harmonic condition respectively (every rising, lowering of indices and operators is done with Minkowski's metric).

The general solution of 
this system can be splitted into a particular solution of the inhomogeneous system plus the general solution of its homogeneous part, which is the same at every order. Therefore, the contribution of the later to the solution up to order $i$ can be written as
\begin{align}
\bm{h}_\text{hom} =&
 \sum_{l=0,2} \lambda \Omega^{l} \frac{m_l}{r_0^l}\, r^l\left(\bm{T}_l+\bm{D}_l\right)
+ \sum_{l=1,3} \,\lambda^\frac32\Omega^{l}\frac{j_l}{r_0^l}\, r^l\,\bm{Z}_l\nonumber
\\
&+ \sum_{l=0,2} \lambda \Omega^{l} \frac{a_l}{r_0^l}\, r^l\bm{E}^*_l
+ \lambda\Omega^{2}{b_2}\, \bm{F}_0^* +\mathcal{O}(\Omega^4)\,.
\label{eq12}
\end{align}
This expression requires several definitions and comments. First, it must noted that $m_l,\, j_l$, $a_l$ and $b_2$ are, in general, series in $\lambda$ and $\Omega$ (see CMMR for details). Next,
\begin{equation}
\begin{aligned}
\bm{T}_l &:= P_l(\cos\theta)\,\bm{\omega}^t\otimes\,\bm{\omega}^t \quad (l\geq 0)\,,
\\
\bm{D}_l &:= P_l(\cos\theta)\,\delta_{ij}dx^i{\otimes\,}dx^j \quad (l\geq 0)\,,
\\
\bm{Z}_l &:=
P_l^1(\cos\theta)\,(\bm{\omega}^t\otimes\bm{\omega}^\phi+\bm{\omega}^\phi\otimes\bm{\omega}^t)
\quad (l\geq 1)\,,
\end{aligned}
\label{eq13}
\end{equation}
are spherical harmonic tensors; $\bm{\omega}^t=dt$, $\bm{\omega}^r=dr$, $\bm{\omega}^\theta=r\,d\theta$, $\bm{\omega}^\phi=r\sin\theta\,d\phi$ 
is the Euclidean orthonormal cobasis correspoding to the spherical-like coordinates $\{t,\phi, r, \theta\}$ associated to the Cartesian-like harmonic coordinates $\{x^\alpha\}$ and $P_l(\cos\theta),\,P^1_l(\cos\theta)$ are associated Legendre polynomials.
  The tensors
\begin{equation}
\begin{aligned}
 \bm{E}^*_l &:=\frac{1+l}{6}\left[(6+4l)\bm{D}_l-l \bm{H}_l\right]- \frac{1}{2}(\bm{H}_l^1+\bm{H}_l^2)\\
\bm{F}^*_l &:=\frac{1}{2}(l+1)(l+2)\bm{H}_l-(l+2)\bm{H}_l^1-\frac{1}{2}\bm{H}_l^2
\label{eq14}
\end{aligned} \end{equation} 
are convenient combinations of $\bm{D}_l$ and these other three spherical harmonic tensors\footnote{Note that the definition of $\bm{E}_l^*$ is different from the one used in \cite{cabezas2007ags}. Also, $\bm{E}_0^*:=\bm{D_0}$ has now  spherical symmetry and the constant $b_0$ has been renamed $b_2$.
}
\begin{equation}
\begin{aligned}
 \bm{H}_l &:= P_l(\cos\theta)\,(\delta_{ij}- 3e_i  e_j)dx^i{\otimes\,}dx^j \quad (l\geq 0)\,,
 \\
\bm{H}^1_l &:= P_l^1(\cos\theta)\,(k_ie_j + k_je_i)dx^i{\otimes\,} dx^j \quad (l\geq 1)\,,
 \\
\bm{H}^2_l &:= P_l^2(\cos\theta)\,(k_ik_j -m_im_j)dx^i{\otimes\,} dx^j \quad (l\geq 2)\,,
\end{aligned}
\label{eq15}
\end{equation}
where $k_i$, $e_i$ and $m_i$ stand for Euclidean unit vectors of the standard cylindrical coordinates, 
$d\rho=k_i\,dx^i$, $dz = e_i\,dx^i$, $\rho\,d\phi = m_i\,dx^i$. 
The scalar coefficients of the tensors in \eqref{eq12} show a $\Omega$ dependance that we chose in the following way.
The base dependence in $\Omega$ is equal to the degree $l$ of the corresponding tensor, which comes from the associated Legendre polynomials in their definition.  An extra $\Omega$ factor can be added to ensure that only tensors with spherical symmetry  be present when $\Omega\rightarrow0$ (e.g. $\bm{F}^*_0$ is not spherically symmetric despite having $l=0$).  Also, it can be shown that $a_0,\,a_2$ and $b_2$ only parametrize changes of coordinates.
Lastly, the equatorial symmetry imposes that every summation runs only over even index $l$, except the one for $\bm{Z}_l$ which runs over odd index. We cut the expansion at $\mathcal{O}(\Omega^3)$. 

The first order  solution of \eqref{eq11} corresponds to the constant energy density problem already solved in \cite{cabezas2007ags}. It is so because
from \cref{eq5,eq6} and the definition of $\lambda$, the first terms of the energy density and pressure are $\epsilon\sim\epsilon_0\sim{\mathcal{O}}(\lambda)$ and $p\sim{\mathcal{O}}(\lambda)^2$. This dependence on $\lambda$ of $\epsilon$ and $p$ is not an assumption but a consequence of the first term of the expression of $\psi$, i.\,e. $\psi=1+\mathcal{O}(\lambda,\,\Omega^2)$, which in turn comes from the post-Minkowskian expansion of $\bm{g}^-$ in harmonic coordinates and the parameter choice \eqref{omegadef} as well as the fact that we choose $\bm{\xi}^-$ to be a unit time translation at Minkowskian level. 
 Besides, a relation of this kind also holds for spherical configurations in Newtonian theory \citep{bradley2009quad}.

Finally, although working with the $\bm{E}_l^*,\,\bm{F}_l^*$ tensors is useful while obtaining the interior, its final expression is more compact when given in terms of \footnote{Note that the definitions from \cite{cabezas2007ags} have been modified. Now $\bm{E}_2$ has spherical symmetry.}
\begin{equation}
\begin{aligned}
&\bm{E}_l := \frac12l(l-1)\,\bm{H}_l+(l-1)\,\bm{H}^1_l -\frac12\,\bm{H}^2_l \quad (l\geq 2)\\
&\bm{F}_l := \frac13l(2l-1)\,\bm{D}_l -\frac16l(l+1)\,\bm{H}_l-\frac12\,(\bm{H}^1_l+\bm{H}^2_l)  \quad (l\geq 1)\,,\end{aligned}
\label{eq21}
\end{equation}
and it also makes the matching procedure faster, so, using these and defining $\eta:={r/r_0}$, the interior solution is
\begin{multline}
\bm{g}^-= -{{\bm{T}_0}+{\bm{D}_0}}+\lambda  \bigg\{ \left({m_0}-\eta ^2\right){\bm{T}_0}+ \left({a_0}+{m_0}-\eta ^2\right){\bm{D}_0}+\vphantom{\frac{1}{2}}\\
\shoveleft{
	\hspace{4em}+\Omega ^2\left[{m_2}  \eta ^2{\bm{T}_2}+\left(5
{a_2} +{m_2} \right)\eta ^2{\bm{D}_2}+{b_2} {\bm{H}_0}+{a_2}  \eta ^2{\bm{F}_2} \right] \bigg\} }\\
\shoveleft{
	\hspace{2em}{}+\lambda ^2 \eta ^2
	\left\{    \left(-{a_0}  -\frac{m_0}{2}(n+2) +(n  +2) S +\left(2+3 n\right)\frac{\eta ^2}{20}\right){\bm{T}_0}     \right.}\\
\shoveleft{
	\hspace{4em}{}+ \left(-2 {a_0}  -\frac{m_0}{2}(n+2) +(n-2)S +\left(\frac{13}{3}+\frac{3n }{2}\right)\frac{\eta ^2}{10}\right){\bm{D}_0}}\\
\shoveleft{
	\hspace{4em}{}+ \left(-\frac{{m_0} }{5}+\frac{2 S }{5}+\frac{2 \eta ^2}{21}\right){\bm{E}_2} }\\
\shoveleft{
	\hspace{4em}{}+\Omega ^2\left[    \left(-\frac{3 }{5}-\frac{n }{10}\right)\eta ^2{\bm{T}_0}+ \left(\frac{1}{15}-\frac{n }{10}\right)\eta ^2{\bm{D}_0}   \right.}\\
\shoveleft{
	\hspace{4em}{}+ \left(\frac{6 }{7}-3 {a_2} +\frac{{m_2} }{7}+\frac{n }{7}-\frac{3}{14} {m_2} n
\right)\eta ^2{\bm{T}_2}    }\\
\shoveleft{
	\hspace{4em}{}+ \left(\frac{2 }{7}-8 {a_2} -{m_2} +\frac{n }{7}-\frac{3}{14} {m_2} n \right)\eta ^2{\bm{D}_2}}\\
\shoveleft{
	\hspace{4em}{}- \left({b_2} +(1+m_2)\frac{\eta ^2}{15}\right){\bm{H}_0}-\frac{2
 \eta ^2}{21}{\bm{E}_2}+\left(\frac13-\frac{m_2}{2}\right) \frac{\eta ^2}{105}{\bm{E}_4}}\\
\shoveleft{
\hspace{4em}{}-\left.\left.       \left(\frac{m_2}{5}(2m_0+a_0) +\left(a_2+\frac{2}{21}(2-m_2)\right)\eta ^2\right){\bm{F}_2}       \right] \right\}
}\\
\shoveleft{
  \hspace{2em}{}+\lambda ^{3/2}\Omega \eta \left[ \left({j_1}  -\frac{6 \eta ^2}{5}\right){\bm{Z}_1}  +\Omega^2{j_3}  \eta ^2{\bm{Z}_3} \right]}\\
\shoveleft{	
\hspace{2em}{}+\lambda ^{5/2} \Omega\eta^3
	\frac15\left\{ \left(j_1-12 (a_0+ m_0)-3m_0 n +6 n S +\left(\frac{27}{7}+\frac{15n}{14}\right)\eta^2\vphantom{\frac{A^{\frac23}}{A^2}}\right){\bm{Z}_1} \right.}\\
\shoveleft{
	\hspace{4em}{}+\Omega^2\left[ -\left(\vphantom{\frac{A^{\frac23}}{A^2}}\frac{42 {b_2} }{5}+{j_1} {m_2} +\left(4-3
m_2-\frac{n m_2}{2}+2 n -\frac{42 a_2 }{5}\right)\frac{3 \eta ^2}{7}\right){\bm{Z}_1}\right.}\\
\shoveleft
{\hspace{4em}{}+\left.\left.   \left(\frac{4 }{9}-\frac{48 {a_2} }{5}+\frac{{5j_3} }{9}-\frac{4 {m_2} }{3}+\frac{2
n }{9}-\frac{{m_2} n}{3}  \right)\eta ^2{\bm{Z}_3}    \right] \right\}}\\
 +\mathcal{O}(\lambda^3,\Omega^4)\hspace*{27em}
\label{metInt}
\end{multline}
where the constant $S$ comes from the expansion of $\psi$ on the surface, $\psi_\Sigma=1+S\lambda+{\mathcal{O}}(\lambda^2)$, and takes the value
\begin{equation}
 S= \frac{m_0-1}{2}  +  \frac {\Omega ^{2}}{3} +\mathcal{O}(\Omega^4) .
 \end{equation}
\section{Global Solution}
\subsection{Exterior Solution}
The approach in $\mathcal V^+$ is entirely similar. We have to solve the  system \eqref{eq11} for vacuum, i.\,e., with $\mathcal{T_{\alpha\beta}}=0$. Its general, homogeneous, regular at infinity solution is 
\begin{multline}
\bm{h}_\text{hom} =
 2\sum_{l=0,2} \lambda\Omega^lr_0^{l+1}\frac{M_l}{ r^{l+1}}\left(\bm{T}_l+\bm{D}_l\right)
+ 2\sum_{l=1,3}\,\lambda^\frac32\Omega^lr_0^{l+1}\frac{J_l}{r^{l+1}}\,\bm{Z}_l
\\
+
\sum_{l=0,2} \lambda\Omega^lr_0^{l+3}\frac{A_{l}}{ r^{l+3}}\bm{E}_{l+2}
+ \lambda\Omega^2r_0^3\frac{B_2}{ r^{3}}\,\bm{F}_2 +\mathcal{O}(\Omega^4)\,.
\label{eq20}
\end{multline}
Here $\bm{T}_l,\,\bm{D}_l\text{ and }\bm{Z}_l$ are defined as in \eqref{eq13}, and $\bm{E}_l,\,\bm{F}_l$ were defined in \eqref{eq21}.
  The base dependence on $\Omega$ of the scalar coefficients of the harmonic tensors has been assigned following the rules already stated for the interior solution. Its important to note nevertheless that, since we work in spherical coordinates associated to harmonic Cartesian-like ones,  $\tilde{M}_l=\lambda\Omega^lr_0^{l+1}{M_l}$ and $\tilde{J}_l=\lambda^\frac32\Omega^lr_0^{l+1}J_l$ are the Thorne-Geroch-Hansen multipole moments \cite{thorne1980meg,geroch1970multipole,hansen1974multipole} and $A_n$ and $B_n$ are constants  parametrizing changes within this family of coordinates. Notice also that
the coordinates have been named as the ones used in $\cal V^-$ but they are not the same in principle. Even after the matching, the coordinates will only be $\mathcal{C}^0$ on the surface. Properly speaking, $\{t,\phi, r, \theta\}$ should be replaced by $\{T, \Phi, R, \Theta\}$ by now.

The general vacuum solution is
\begin{align}
 \bm{g}^+&=  -{\bm{T}_0}+{\bm{D}_0}+\lambda\frac1\eta  \left[2  {M_0} ({\bm{T}_0}+{\bm{D}_0})+\frac{{A_0}  }{{\eta}^2}{\bm{E}_2}\right.\nonumber
\\
&\hspace{8.5em}\left. +\,\Omega ^2\frac{1}{\eta^2}\left(2  {M_2}({\bm{T}_2}+{\bm{D}_2})+{B_2}{\bm{F}_2} +\frac{{A_2}  }{{\eta}^2}{\bm{E}_4}\right)\right]\nonumber
 \\
&
\quad+\lambda ^2\frac{1}{\eta^2} \left\{
\left(-2M_0^2 -\frac{{A_0} {M_0} }{{\eta}^2}\right) {\bm{T}_0}+\left(\frac{4 M_0^2 }{3}-\frac{{A_0} {M_0} }{{\eta}^2}+\frac{3 {A_0}^2 }{2{\eta}^4 }\right){\bm{D}_0}\right.\nonumber
\\
&
	\hspace{4em}+ \left(-\frac{{M_0}^2 }{3 }+\frac{2 {A_0} {M_0} }{{\eta}^2}-\frac{9 {A_0}^2 }{4{\eta}^4
}\right)   {\bm{E}_2}  \nonumber
\\
&\hspace{4em}+\Omega ^2\frac{1}{\eta^2}\left[ 
 {A_0}\left(\frac{6  {B_2} }{5{\eta}^2 }+\frac{2 
{M_2} }{5 {\eta}^2}-\frac{3  {A_2} }{2 {\eta}^4}\right)\bm{H}_{0}
\right.\nonumber
\\ 
&\hspace{4em}+\,\left(2 M_0( B_2 -2 M_2)-\frac{3 {A_2} {M_0}
}{{\eta}^2}-\frac{3 {A_0} {M_2} }{{\eta}^2}\right) {\bm{T}_2}\nonumber
\\
&+
\left(2 {M_0}\left({B_2}+\frac{32 {M_2} }{21 }\right)-\frac{A_0}{7{\eta}^2}\left(12  {B_2}-13  {M_2} \right)-\frac{3 {A_2} {M_0} }{{\eta}^2 }+\frac{66 {A_0} {A_2} }{7 {\eta}^4}\right){\bm{D}_2}\nonumber
\\
&\hspace{4em} +\left(2M_0\left( {B_2} -\frac{2 {M_2}}{21  }\right)-\frac{A_0}{7{\eta}^2}\left(15  {B_2}  -4  {M_2} \right)+\frac{30 {A_0} {A_2} }{7 {\eta}^4}\right){\bm{F}_2}\nonumber
\\
&
\hspace{4em}\!\left.\left.+ \left(-\frac{{M_0} {M_2} }{7 }+\frac{A_0}{35{\eta}^2}\left(33 {B_2} +6  {M_2} \right)+\frac{2 {A_2} {M_0}
}{{\eta}^2}-\frac{53 {A_0} {A_2} }{14 {\eta}^4 }\right){\bm{E}_4}
\right] \right\}\nonumber  
\\
&\quad+\lambda ^{3/2}\Omega \frac{2}{\eta^2}\left( J_1  {\bm{Z}_1} +\Omega ^2\frac{{J_3}  }{{\eta}^2}{\bm{Z}_3} \right)
+\lambda ^{5/2}\Omega\frac{1}{{\eta}^3}
\left\{ \left(-2 {J_1} {M_0} -\frac{{A_0} {J_1} }{{\eta}^2}\right) {\bm{Z}_1}  \right.\nonumber
\\
&
\hspace{4em}+\Omega ^2\frac{1}{{\eta}^2}\left[\left(\frac{4 {B_2} {J_1} }{5 }+\frac{9 {A_2} {J_1} }{5 \eta^2}\right) {\bm{Z}_1}\right.\nonumber
\\
&\hspace{4em}\left.\left.+\left(\frac{6 {B_2} {J_1} }{5 }-{J_3} {M_0} -{J_1} {M_2} +\frac{{A_2}
{J_1} }{5 {\eta}^2}-\frac{3 {A_0} {J_3} }{{\eta}^2}\right){\bm{Z}_3}\right] \right\}
+\mathcal{O}(\lambda^3,\Omega^4)
\label{metExt}
\end{align}

\subsection{Matching}

The last step to build our global spacetime $\mathcal{M}$ is the matching of $({\cal V}^-,\bm{g}^-)$ and $({\cal V}^+,\bm{g}^+)$. It requires first a point to point identification of a hypersurface $\Sigma^-$ of ${\cal V}^-$ and another one $\Sigma^+$ of ${\cal V}^+$. Later, we need  to impose some conditions on $\bm{g}^-$ and $\bm{g}^+$ on these surfaces in order to ensure a reasonably smooth behavior of both metrics considered as parts of a solution of Einstein's equations on $\mathcal{M}$ (matching conditions). The above mentioned identification between hypersurfaces can be implemented by looking at $\Sigma^-$ and $\Sigma^+$ as the embeddings into ${\cal V}^-$ and  ${\cal V}^+$ of a three dimensional manifold $\Sigma$, to which we refer hereafter as the matching surface. Even though several matching conditions have been used in the literature \cite{bonnor1981junction}, it is widely accepted that those proposed by Darmois are the more general and suitable to the matching problem \cite{darmois1966memorial}. 
They impose the first and second 
fundamental forms of $\Sigma^-$ and $\Sigma^+$ to be equal, i.\,e.\!\! continuous through the matching surface.
In particular, these conditions imply that in the neighborhood of every point of  $\Sigma$ there is a local set of coordinates in which the metric and their first derivatives are continuous  \cite{lichnerowicz1955theories}. These coordinates are called  \textsl{Lichnerowicz admissible coordinates} and to require the metric to be of class $\mathcal{C}^1$ on the matching surface is known as  \textsl{Lichnerowicz matching}.

Darmois and Lichnerowicz matching conditions are equivalent in the sense that the latter are the practical realization of the former in a certain set of coordinates. Nevertheless, if one fails to match two metrics written in some definite coordinate systems using Lichnerowicz conditions, it does not imply that the metrics can not be indeed matched.

\subsubsection{Matching surface}

Here, we are going to match the exterior and interior solutions given in the previous sections keeping all the free constants they have. First, let us start by discussing how to choose the matching surface. In a general matching of spacetimes the surface, along with its embeddings, is actually part of the solution and can not be given beforehand without risk of losing generality (see \cite{mars2007linear} and \cite{mars1998cgm}  for a discussion of this topic). In our case and in stellar model building, $\Sigma^-$ is uniquely  characterized in $\mathcal{V}^-$ as the locus of points where $p=0$ and it is the only relevant matching surface.  The zero pressure surface $r=r_\Sigma (\theta)$ of our interior metric is defined implicitly by the equation $\psi(r,\theta)=\psi_\Sigma$ (see \cref{eq6}). Therefore, we may write
\begin{equation}
 \Sigma^-=\{t=\tau,\,\phi=\varphi,\,r=r_\Sigma(\vartheta),\, \theta=\vartheta)\}.
\end{equation} 
where $\{\tau,\,\varphi,\,\vartheta\}$ are coordinates of $\Sigma$.

To simplify the resolution of the matching, we are interested 
 in using a common expression for $\Sigma^+$ and $\Sigma^-$ in terms of the interior an exterior coordinates, that is to say
 \begin{equation}
 \Sigma^+=\{T=\tau,\,\Phi=\varphi,\,R_\Sigma=r_\Sigma(\vartheta),\, \Theta_\Sigma=\vartheta)\}\,,
 \label{surfacemas}
\end{equation} 
$\{T, \Phi, R, \Theta\}$ being coordinates in ${\cal V}^+$. This assumption implies no loss of generality if we prove that the coordinates that verify the above equations for $\Sigma^-$ and $\Sigma^+$ belong to the class of coordinates we use to write the metrics. Then we can go along with the matching using the expressions for the interior and exterior metrics we got in the previous sections.

Let us set up the problem. The following expression for $\Sigma^+$,
\begin{equation}
 \Sigma^+=\{T=\tau,\,\Phi=\varphi,\,R=R(\vartheta),\, \Theta=\Theta(\vartheta)\}.
\end{equation} 
is well suited to the symmetries of the exterior field and how they have been implemented in the metric \eqref{metExt}. Coordinates $T$ and $\Phi$ are adapted to the Killing fields, and they have been chosen to ensure that the metric tends to the flat metric in standard spherical coordinates at infinity. They are unique up to an additive constant we can set equal to zero. On the other hand, coordinates $R$ and $\Theta$ are not completely set. Any pair of functions $F(R,\Theta)$ and $H(R,\Theta)$ leading to a set of Cartesian-like harmonic coordinates $X'=F(R,\Theta)\cos\Phi$, $Y'=F(R,\Theta)\sin\Phi$ and $Z=H(R,\Theta)$ defines implicitly a couple of new coordinates by means of these two equations $R'\cos\Theta'=F(R,\Theta)$ and $R'\cos\Theta'=H(R,\Theta)$. Nevertheless, we must impose some conditions on the two functions in order to preserve the good behavior of the coordinates at infinity, namely
\begin{equation}
F(R,\Theta)\rightarrow R\sin\Theta\,,\quad
H(R,\Theta)\rightarrow R\cos\Theta\qquad
(R\rightarrow\infty)
\end{equation}
This freedom is actually included in our metric \eqref{metExt} by means of the constants $A_0$, $A_2$ and $B_2$.

The harmonic condition requires $F$ (and $H$) to be a solution of a second order elliptic equation. If we add to the boundary condition at infinity  mentioned above  this other one on $\Sigma^+$,
\begin{equation}
F\left((R_\Sigma(\vartheta),\Theta_\Sigma(\vartheta)\right)=r_\Sigma(\vartheta)\sin\vartheta\,,\quad
H\left(R_\Sigma(\vartheta),\Theta_\Sigma(\vartheta)\right)=r_\Sigma(\vartheta)\cos\vartheta\,,
\end{equation}
we get a Dirichlet problem. We assume by now that it admits a solution. We will later show that this is indeed the case, at least up to the order considered, since we are able to find a solution of the  Lichnerowicz matching in these coordinates.
 The equation of the surface $\Sigma^+$ in the new coordinates (we drop out the primes) can then take the form we want, \cref{surfacemas}.  

There is another problem which should be mentioned here, even though it has no consequences on the equation for the matching surface itself. It has been pointed out that coordinates $T$ and $\Phi$ can not be naively identified with $t$ and $\phi$ on the matching surface as we have done in the precedent paragraph  \cite{mars1998cgm}. Anyhow it can be done by making a suitable linear change, $t=at'$ and $\phi=\phi'+abt'$ preserving the regularity of the symmetry axis.

In order to make such a change compatible with the approximate interior metric (\cref{metInt}), we have to assume an expansion of these two constants in powers of $\lambda$ as follows,  $a=1+\mathcal{O}(\lambda)$ and $b=\mathcal{O}(\lambda^{1/2})$. The first and the lack of a $\mathcal{O}(\lambda^0)$ term in $b$ just take into account that the starting point of our approximation is the flat metric; the reason behind the semi-integer expansion of $b$ is that under this change the angular velocity of the fluid reads $\omega-b$, and within the CMMR scheme this quantity must be of order $\mathcal{O}(\lambda^{1/2})$. These infinitesimal expansions actually do not keep the structure of the interior metric because $g^-_{t'\phi'}$ contains a $\mathcal{O}(\lambda^{1/2})$ term proportional to $b$ which is absent in $g^-_{t\phi}$ (that starts at $\mathcal{O}(\lambda^{3/2})$). It does not matter. The matching sets it equal to zero because the component $g^+_{T\Phi}$ can not have a $\mathcal{O}(\lambda^{1/2})$ term 
unless it violates the asymptotic conditions on the coordinates. Moreover, the rest of the 
contribution of $a$ and $b$ to the metric in the new coordinates can be absorbed into the free constants $m_0$, $a_0$ and $j_1$.

We actually think that the choice of $t$ and $\phi$ discussed above has already been considered in the CMMR scheme from the very beginning. Dropping out a $\mathcal{O}(\lambda^{1/2})$ in $g^-_{t\phi}$ at the linear level of the approximation is a way of choosing the inner coordinates $t$ and $\phi$, and also the angular velocity $\omega$ in a sense, since we asumed $g^+_{T\Phi}$ to be proportional to $\omega$ in \cref{eq10}.

These kind of assumptions are meaningful in a perturbation scheme but they can not easily be implemented in a exact matching problem. The continuity of the Killing fields on the surface matching used in  \cite{mars1998cgm}  is a smart reasonable assumption to make an exhaustive use of the symmetries of the problem. The arguments sketched above show how this point of view has not been forgotten in our scheme. Therefore, we can argue that our approach is in accordance with it.

Lastly, let us simplify a little bit more the matching process by introducing an ansatz for the explicit equation of the surface. Being this an axisymmetric problem, the zero pressure surface can be expanded as a power series in Legendre polynomials, and the reflection symmetry allows us to rule out every odd degree term in the expansion. Coherently with the assignment of dependence on $\Omega$ we made for the approximate solutions of the homogeneous linear Einstein equations, we have then
\begin{equation}
 r_\Sigma(\theta)= r_0\left\{1+\Omega^2\left[\sigma_0 + \sigma_2  P_2 (\cos\theta)\right]\right\}+{\cal O}(\Omega^4)
\label{eq7}
\end{equation}
where $r_0$ is the coordinate radius of the fluid at rest and $\sigma_{0,2}$ are constants expandable in $(\lambda,\,\Omega)$ to be determined while matching. This is also coherent with the kind of expression we need for $r$ to solve the implicit equation $\psi(r,\,\theta)=\psi_\Sigma$ of the surface.
\subsubsection{Darmois matching}
%
%
%
%
%

First we impose Darmois conditions 
 on the surface \eqref{eq7} to match $\{\mathcal{V^-},\bm{g}^-\}$ and $\{\mathcal{V^+},\bm{g}^+\}$ up to $\mathcal{O}(\lambda^2,\Omega^3)$ in the metrics. They are satisfied when the constants associated to the multipole moments are
\begin{align}
M_0&= 1+\lambda  \left[\frac{3 a_0}{2}+\frac{n}{5}+\frac{14}{5}+\frac{2}{15}\Omega ^2\left(4- n\right) \right]+\mathcal{O}(\lambda^2,\Omega^4),
\label{mass1}\\
M_2&= -\frac{1}{2}+\lambda  \left(-\frac{5 a_0}{4}+\frac{n}{14}-\frac{37}{35}\right)+\mathcal{O}(\lambda^2,\Omega^2),
\label{mass2}\\
J_1&= \frac{2}{5}+\frac{\Omega ^2}{3}+ \lambda  \left[a_0+\frac{2 n}{35}+\frac{16}{7}+\Omega ^2 \left(\frac{5 a_0}{6}-\frac{3 n}{35}+\frac{176}{105}\right)\right]+\mathcal{O}(\lambda^2,\Omega^4),
\label{angular1}
\\
J_3&= -\frac{1}{7}+\lambda  \left(-\frac{a_0}{2}+\frac{11 n}{441}-\frac{496}{735}\right)+\mathcal{O}(\lambda^2,\Omega^2),
\label{angular2}
\intertext{while their interior counterparts take the values}
m_0&=3+ \lambda  \left[3 a_0+\frac{3 n}{4}+\frac{9}{2}+\Omega ^2\left(1-\frac{n}{2}\right) \right]+\mathcal{O}(\lambda^2,\Omega^4),
\label{int1}\\
m_2&= -1+\lambda  \left(-a_0+2 b_2-\frac{3 n}{14}-\frac{29}{35}\right)+\mathcal{O}(\lambda^2,\Omega^2),
\label{int2}
\\
j_1&= 2+\frac{2 \Omega ^2}{3}+\lambda  \left[4 a_0+\frac{n}{2}+\frac{49}{5}+\Omega ^2 \left(\frac{4 a_0}{3}+2 b_2-\frac{5 n}{14}+\frac{289}{105}\right)\right]+\mathcal{O}(\lambda^2,\Omega^4),
\label{int3}\\
j_3&= -\frac{2}{7}+\lambda  \left(-\frac{4 a_0}{7}+2 a_2+\frac{12 b_2}{25}-\frac{3 n}{49}-\frac{326}{245}\right)+\mathcal{O}(\lambda^2,\Omega^2)
\label{int4}
\intertext{and finally, the exterior gauge constants are}
A_0&= a_0+\lambda  \left(\frac{3 a_0^2}{4}-3 a_0-\frac{237}{35}-\frac{22 \Omega ^2}{35}\right)+\mathcal{O}(\lambda^2,\Omega^4),
\label{gauge1}\\
A_2&= -\frac{a_0}{2}+a_2+\lambda  \left(\frac{a_0 n}{7}-\frac{7 a_0^2}{8}+a_2 a_0+\frac{139 a_0}{70}-3 a_2+\frac{383}{90}\right)+\mathcal{O}(\lambda^2,\Omega^2),
\label{gauge2}
\\
B_2&= \frac{a_0}{2}+b_2+\lambda  \left(-\frac{a_0 n}{7}+\frac{3 a_0^2}{8}-\frac{25 a_0}{14}-3 b_2-\frac{7}{2}\right)+\mathcal{O}(\lambda^2,\Omega^2)\,.
\label{gauge3}
\end{align}
The matching surface on which the matching conditions hold is
\begin{equation}
\eta_\Sigma=1+P_2(\cos\theta) \Omega ^2 \left[-\frac{5}{6}+\lambda  \left(-\frac{3
   a_2}{2}+b_2+\frac{5 n}{21}+\frac{10}{21}\right)
\right]
+\mathcal{O}(\lambda^2,\Omega^4).
\end{equation} 
These results require some comments. Here, the only free parameters are $a_0,\,a_2$ and $b_2$, (which, as already mentioned, parametrize changes of harmonic coordinates in $\mathcal{V}^-$), $r_0$ (which depends on the size of the source \textit{but also on the coordinates}), $\epsilon_0$, $\omega$ (which are part of the definitions of $\lambda$ and $\Omega$) and the EOS parameter $n$. Then, for a fixed set of source parameters $s:=\{n,\,\epsilon_0,\,\omega\}$ and $r_0$,\footnote{Note that $r_0$ has been excluded from $s$ because of its coordinate dependence. It contains information necessary to fully characterise the source, though.} the interior 
metric that can be matched is unique up to changes of coordinates. Since the asymptotically flat exterior of a certain source spacetime in rotation  is unique \cite{mars1998cgm} then, given a set of source parameters $s$ and $r_0$, there is only one possible global spacetime, i.\,e., only a couple of metrics $\left( \hat{\bm{g}}^-,\,\hat{\bm{g}}^+ \right)$ among the families $\bm{g}^-$ and $\bm{g}^+$ give 
 spacetimes that can be matched. Nevertheless, this could seem contradictory with the apparent fact that the value of mass and angular multipole moments depend on the value of $a_0
 $ as \cref{mass1,mass2,angular1,angular2}. This apparent dependence happens because these constants are not the only gauge dependent quantities in the expressions. There is coordinate dependence hidden in $\lambda$, which, unlike $\Omega$, depends on $r_0$.

This problem can be solved finding their expression in terms of physical constants because they are gauge invariant. A convenient way of doing it is using the  mass monopole moment $\tilde{M}_0$ to redefine $\lambda$. In CMMR, $\lambda$ is defined as $\lambda:=\frac{4\pi}{3}\epsilon_0 r_0^2$ so that 
\begin{equation}
 \tilde{M}_0=-\lim_{r\rightarrow\infty}\frac12g^+_{tt,r}r^2=\lambda r_0 M_0
\end{equation} 
and $M_0=1+\mathcal{O}(\lambda)$ reflecting the fact that $\lambda$ and $r_0$  were chosen to reproduce the Newtonian mass of the source. 
 Now, if we define $\{\lambda',r_0'\}$ to give the general relativistic mass monopole moment $\tilde{M}_0$ and keep the same relation between them, i.\,e.
\begin{align}
 \lambda'&:=\frac{4\pi}{3}\epsilon_0 {r'_0}^2,\label{lambdas1}\\
 \lambda' r'_0&:=\tilde{M}_0\label{lambdas2},
\end{align}
inserting \cref{lambdas1} into \cref{lambdas2} and using the expression of $M_0$ \eqref{mass1} to find $\tilde{M}_0=\lambda r_0 M_0$, we obtain that
\begin{align}
 \lambda'&=\lambda\left(1+\lambda \left\{a_0+\frac23\left[\frac{n}{5}+\frac{14}{5}+\frac{2}{15}\Omega^2(4-n)\right]\right\}\right)+\mathcal{O}(\lambda^3,\Omega^4),\\
r_0'&=r_0\left(1+\lambda\left\{\frac{a_0}{2}+\frac13\left[\frac{n}{5}+\frac{14}{5}+\frac{2}{15}\Omega^2(4-n)\right]\right\}\right)+\mathcal{O}(\lambda^2,\Omega^4).
\end{align}
Using these changes, $M_0=1+\mathcal{O}(\lambda'^2)$ as required and the  dependence on the gauge constant $a_0$ of the multipole moments $\tilde{M}_2,\,\tilde{J}_1$ and $\tilde{J}_3$ dissapears. In this way, the set of parameters to completely specify the interior would become $\epsilon_0,\, n,\, \omega$ and $\tilde{M}_0$. 

This procedure can be followed as well using the central pressure $p_c$ instead of $\tilde{M}_0$ to characterise the interior as is sometimes done  in astrophysics. In this case, the new approximation parameter $\Lambda$ is
\begin{equation}
\Lambda=\lambda\left\{1+\lambda \left[a_0+\frac{3 n}{5}+\frac{17}{5}+\Omega ^2\left(\frac{3}{5}-\frac{n}{6}\right) \right]\right\}+\mathcal{O}(\lambda^3,\Omega^4),
\end{equation} 
with
\begin{equation}
\Lambda:=p_c \frac{2}{\epsilon _0}\left(1+\frac23 \Omega ^2\right) +\mathcal{O}(\Omega^4)   .                                                                           \end{equation} 
Keeping a relation of the form of \cref{lambdas1}, the associated radial coordinate changes as 
\begin{equation}
 r_\Lambda=r_{0}\left\{1+\lambda  \left[\frac{a_0}{2}+\frac{3 n}{10}+\frac{17}{10}+\Omega ^2\left(\frac{3}{10}-\frac{n}{12}\right) \right]\right\}+\mathcal{O}(\lambda^2,\Omega^4).
\label{paramchanger0}
\end{equation}
Again, as one expects, with these changes the expressions for the multipole moments become manifestly coordinate independent.

 Thus, with these last results, the exterior \eqref{metExt} and the interior \eqref{metInt} with their constants taking the values \cref{mass1,mass2,angular1,angular2,int1,int2,int3,int4,gauge1,gauge2,gauge3} give the most general approximate family of global asymptotically flat solutions for the kind of source studied, each of its members characterized only by the values of $\{n,\,\epsilon_0,\,\omega\}$ and $r'_0$ or $r_\Lambda$ that, with the first three fixed, depend only on $\tilde{M}_0$ and $p_c$, respectively.  Additionally, they  also point out the behaviour one intuitively  expects as a generalization of the theorem in \citep{rendall1991existence}, i.\,e., that for a stationary axisymmetric singularity free compact rotating perfect fluid, its asymptotically flat exterior is unique once the EOS, central pressure and rotation speed are fixed.
\subsubsection{Lichnerowicz matching}

Now we impose Lichnerowicz conditions 
. The $\mathcal{O}(\lambda^2,\Omega^3)$ metric is then matched in the global set of coordinates when its multipole moments and $\{m_l,\,j_l\}$ cons\-tants take the values
\begin{align}
 M_0&=1+\lambda  \left[\frac{n}{5}+\frac{14}{5}+\Omega ^2\left(\frac{8}{15}-\frac{2 n}{15}\right) \right]+\mathcal{O}(\lambda^2,\Omega^4),\label{M0}\\
M_2&=-\frac{1}{2}+\lambda  \left(\frac{n}{14}-\frac{37}{35}\right)+\mathcal{O}(\lambda^2,\Omega^2),\label{M2}
\\
 J_1&=\frac{2}{5}+\frac{\Omega ^2}{3}+\lambda  \left[\frac{2 n}{35}+\frac{16}{7}+\Omega ^2\left(\frac{176}{105}-\frac{3 n}{35}\right) \right]+\mathcal{O}(\lambda^2,\Omega^4),\label{J1}\\
J_3&=-\frac{1}{7}+\lambda  \left(\frac{11 n}{441}-\frac{496}{735}\right)+\mathcal{O}(\lambda^2,\Omega^2),\label{J3}
\\
 m_0&=3+\lambda  \left[\frac{3 n}{4}+\frac{9}{2}+\Omega ^2\left(1-\frac{n}{2}\right) \right]+\mathcal{O}(\lambda^2,\Omega^4),\label{m0}\\
m_2&=-1+\lambda  \left(-\frac{3 n}{14}-\frac{29}{35}\right)+\mathcal{O}(\lambda^2,\Omega^2),\label{m2}
\\
 j_1&=2+\frac{2 \Omega ^2}{3}+\lambda  \left[\frac{n}{2}+\frac{49}{5}+\Omega ^2\left(\frac{289}{105}-\frac{5 n}{14}\right) \right]+\mathcal{O}(\lambda^2,\Omega^4),\label{j1}\\
j_3&=-\frac{2}{7}+\lambda  \left(-\frac{3 n}{49}-\frac{326}{245}\right)+\mathcal{O}(\lambda^2,\Omega^2),
\label{lich1}
\end{align}
and the coordinate-parametrizing constants are
\begin{equation}
\begin{aligned}
 A_0&=\frac{4\lambda}{35}  \left(2+\frac{ \Omega ^2}{3}\right)+\mathcal{O}(\lambda^2,\Omega^4),\\
A_2&=-\frac{4 \lambda }{63}+\mathcal{O}(\lambda^2,\Omega^2),\\
B_2&=\mathcal{O}(\lambda^2,\Omega^2),
\end{aligned}\qquad
\begin{aligned}
 a_0&=\lambda  \left(7+\frac{2 \Omega ^2}{3}\right)+\mathcal{O}(\lambda^2,\Omega^4),\\
a_2&=-\frac{86 \lambda }{105}+\mathcal{O}(\lambda^2,\Omega^2),\\
b_2&=\mathcal{O}(\lambda^2,\Omega^2).
\end{aligned}
\label{lich2}
\end{equation} 

The value of these parameters is unique for each set of 
 parameters $\{n,\,\epsilon_0,\,\omega,\,r_0\}$ and 
 verify the relations obtained with Darmois conditions as is to be expected. 
 Nevertheless, it proves the existence of a harmonic and asymptotically Cartesian global system of coordinates 
 up to the approximation order considered. 

In spite of the attention drew to Darmois matching before, it is important to remark that the real focus of the approximation scheme is the obtention of totally matched spacetimes in the sense that even the gauge constants are fully fixed and the Lichnerowicz admissible coordinates  are found. In fact, only Lichnerowicz conditions were used in \citep{cabezas2007ags,martin2008crr}, although the generality of the results in them can be verified with the same techniques used here. Besides, one needs the fully matched spacetime for many practical purposes, as for example to compare with numerical results for stellar models \citep{cuchi2012compaProcRevTeX,cuchi2013comparingRevTeX} built using global coordinates.

\section{Petrov classification}
We will now analyse the possible Petrov types of the unmatched interior metric. They are given by the possible Jordan canonical forms of the ${Q^\alpha}_\beta$ matrix, defined from  the Weyl tensor $ {C^\alpha}_{\beta\gamma\delta}$ as
\begin{equation*}
  {Q^\alpha}_\beta:=\left({C^\alpha}_{\lambda\beta\mu}+\text{i} {{^\star C}^\alpha}_{\lambda\beta\mu}\right)u^\lambda u^\mu
\end{equation*}
 where $\star$ denotes the left Hodge dual and $u^\alpha$ is a 
unit
 timelike vector.
 We use $u^i=0$ for simplicity. Due to the symmetries of the Weyl tensor, ${Q^\alpha}_\beta u^\alpha=0$. This implies that for the eigenvalue problem
 \begin{equation}
{Q^\alpha}_\beta v^\alpha=\varepsilon v^\alpha
 \end{equation} 
$v^\alpha u_\alpha=0$ and thus, for a metric with Papapetrou's structure, ${Q^\alpha}_\beta$ is completely classified studying
\begin{equation}
 \bm{Q}=\begin{pmatrix}
  {Q^r}_r	&{Q^r}_\theta	&0\\
  {Q^\theta}_r	&{Q^\theta}_\theta	&0\\
  0		&0		&{Q^\varphi}_\varphi
        \end{pmatrix}\,,
\label{matrizq}
\end{equation} 
that has the greatly simplifying property of possesing two orthogonal blocks, so that we can write all its possible Jordan canonical forms as
\begin{equation}
 \bm{J}_Q=\begin{pmatrix}
  \varepsilon_1	&a	&0\\
  0	&\varepsilon_2	&0\\
  0		&0		&\varepsilon_3
        \end{pmatrix}\,,
\label{matrizJ}
\end{equation}
with $a=0$ or $1$, the latter being a possibility only if $\varepsilon_1=\varepsilon_2$.

 The $\bm{Q}$ matrix is always trace-less and in an orthonormal cobasis also sym\-me\-tric. For computation convenience we work now in spherical-like coordinates and thus ${Q^r}_\theta\neq{Q^\theta}_r$ in general. The eigenvalues $\varepsilon_1$ and $\varepsilon_2$ of the $r-\theta$ subspace are degenerate iff the discriminant of the roots of its characteristic equation is zero
, i.\,e.,
\begin{equation}
 \left( {Q^r}_r-{Q^\theta}_\theta \right)^2+4{Q^r}_\theta {Q^\theta}_r=0
\label{deg1}
\end{equation} 
 which using the traceless property gives 
 \begin{equation}
{Q^r}_r{Q^\theta}_\theta-{Q^r}_\theta{Q^\theta}_r-\frac14\left({Q^\varphi}_\varphi\right)^2=0.
\label{deg1b}
\end{equation} 
The condition for either $\varepsilon_1$ or $\varepsilon_2$ to be degenerate with $\varepsilon_3={Q^\varphi}_\varphi$ is
\begin{equation}
 {Q^r}_r {Q^\theta}_\theta  - {Q^r}_\theta {Q^\theta}_r + 2 ({Q^\varphi}_\varphi)^2=0\,,
\label{deg2}
\end{equation} 
that has the same expression as \cref{deg1b} switching the numerical factor. The structure of $\bm{Q}$, see \eqref{matrizq}, allows the following possibilities.  First, no degeneracy at all. This is the general case and corresponds to Petrov type I, whose Segr\`e symbol is $\{111\}$.\footnote{In a Segr\`e symbol, each number gives the dimension of one of the invariant subspaces. Numbers associated to subspaces with degenerate eigenvalues are written inside parenthesis}
 The second, degeneracy in only two eigenvalues. It can come from
\cref{deg1b}, in which case it can lead to types D --Segr\`e symbol $\{(11)1\}$--  and II --$\{21\}$--, or from \cref{deg2} that can only give rise to type D because from the structure of \eqref{matrizJ} we see that the degenerate eigenvalues would then belong to orthogonal subspaces. Last, three degenerate eigenvalues, what happens if both \cref{deg1b,deg2} are fulfilled. This can lead only to types N --$\{(21)\}$-- and O --$\{(111)\}$-- since type III --$\{3\}$-- is directly excluded again by the form of \eqref{matrizJ}.

We get the following
conditions on the metric parameters
 (in which the trivial case $\lambda=0$ is not considered) extracting from \eqref{deg1b} and \eqref{deg2} 
the relevant information up to this order of approximation.

We will first analyse the general rotating case separately from the static case for clarity. The relevant condition \cref{deg1} for $\varepsilon_1=\varepsilon_2$, even considering only its terms  up to $\mathcal{O}(\lambda^3,\Omega^3)$ 
\begin{multline}
 \smash{-\text{i}\lambda ^{5/2}\Omega ^3\frac{54r
     }{5 r_0^5}  m_2 P_1(\cos\theta)
+\lambda ^3  \Omega
   ^2\frac{3r^2
  }{25 r_0^6}
\times}\\[1.5ex]\times
\left\{5 m_2 (n-1) \left[4 P_2(\cos\theta)-1\right]-108\right\}+\mathcal{O}(\lambda^{7/2},\Omega^4)=0
\label{deg3}
\end{multline} 
can not be satisfied everywhere unless $\Omega=0$ and therefore we can not have $\varepsilon_1=\varepsilon_2$ degeneracy out of the static case.
The conditions for the other possible degeneracies $\varepsilon_1=\varepsilon_3$ or $\varepsilon_2=\varepsilon_3$ are given by \cref{deg2}, that yields
\begin{multline}
 - \lambda ^3  \Omega ^2\frac{3 r^2}{25 r_0^6}  \left[P_2(\cos\theta)-1\right]\left[5 m_2
   (n-1)+18\right]+ \text{i} \lambda ^{7/2}
   \Omega ^3 \frac{18 r^3}{175 r_0^7}\times
\\
\times
  \left[P_1(\cos\theta)-P_3(\cos\theta)\right] \left[2 \left(35 j_3+3\right) (n-1)+m_2
   (23-14 n)\right]
\\
+\mathcal{O}(\lambda^4,\Omega^4)=0.
\end{multline}
 They are satisfied in the static case and, if  $\Omega\neq0$, when the constants of the metric verify 
\begin{equation}
m_2=\frac{18}{5(1-n)}\quad\text{and}\quad j_3=\frac{3(n+8)(8-5n)}{175(n-1)^2}\,,
\label{deg2cond}
\end{equation} 
which can never be satisfied in the constant energy density case.

Concerning the static case, equation \eqref{deg1} gives conditions different from \cref{deg3}. They are satisfied only when $n=1$, while the other degeneracy possibility condition \eqref{deg2} is always verified. This can be seen straightaway from the form the ${\bm{Q}}$ matrix takes 
\begin{equation}
{\bm{Q}}= \frac{1}{5} \frac{\eta ^2}{r_0^2} \lambda ^2\begin{pmatrix}
 2 (n-1) & 0 & 0 \\
 0 &  1-n  & 0 \\
 0 & 0 &  1-n 
\end{pmatrix}+\mathcal{O}(\lambda^3)
\end{equation} 
as well as the fact that the only possible Petrov types are D ($n\neq1$) and O ($n=1$) as must be the case for a spherically symmetric spacetime \cite[p. 228]{stephani2003exact}. The condition on $n$ for type O is expectable since any conformally flat perfect fluid solution with our symmetries must be Schwarzschild's interior solution \cite{collinson1976uniqueness}. From the equations \eqref{deg1} and \eqref{deg2} alone one can not say whether this behaviour will endure further approximations because extra conditions could impose a more general Petrov type, but these theorems on spherical symmetry and constant energy density give a strong hint about its endurance.

Therefore, collecting the results together from the rotating and static cases, we conclude that
\begin{itemize}
\item  An invariant subspace of dimension 3 for the eigenvalue problem, which would lead to Petrov type III, is ruled out by the structure of $\bm{Q}$ associated to the Papapetrou structure of the metric. From our perturbation theory results, we see that the only option for a bidimensional invariant subspace appears in the static limit, where its existence is forbidden by the fact that the spherical symmetry associated imposes types D or O. Accordingly, the $\bm{Q}$ matrix of our interior spacetime is always diagonalizable.

\item In general, the Petrov type is I. Out of the static case, it will only be type D when \cref{deg2cond} are satisfied provided $n\neq1$. In the static case, it will always be type D unless $n=1$, in which case the Petrov type is O. Then, the constant energy density case can only be type I($\Omega\neq0$) or O($\Omega=0$).

It could be argued that our approximate results do not necessarily hold for exact solutions. Nevertheless, as long as an exact solution for the source we work with exists, a series development of it following the CMMR approach must lead to our results. Because of this, while any property compatible with a truncated series development is not necessarily a property of the exact solution, a behaviour ruled out already in the truncated solution can not be a property of the hypothetical exact solution.
\end{itemize}

\section{Some implications}
It has long been suspected that the Kerr metric can not represent the exterior of any stellar model. We can easily check that it is indeed the case here using the kind of analysis that appears in \cite{cabezas2007ags}.

The first three Kerr multipole moments are \cite{hansen1974multipole}\footnote{Here $m$ stands for the Kerr mass parameter.}
\begin{equation}
 {M}^{\text{Kerr}}_0=m,\qquad {J}^{\text{Kerr}}_1=ma,\qquad{M}^{\text{Kerr}}_2=-ma^2.
\end{equation} 
If our two first multipole moments were equal to the Kerr ones,  ${M}^{\text{Kerr}}_2$ should have the expression
\begin{equation*}
\left.
\begin{aligned}
{M}_0^\text{Kerr}&=m=\lambda r_0 M_0\\
{J}_1^\text{Kerr}&=ma=\lambda^{3/2}\Omega r_0^2 J_1
\end{aligned}\right\}\longrightarrow
-ma^2=-\dfrac{(\lambda^{3/2}\Omega r_0^2 J_1)^2}{\lambda r_0 M_0}=-\lambda^2\Omega^2r_0^3\dfrac{ J_1^2}{ M_0}
\end{equation*}
i.\,e., the first $\lambda$-order component of $M_2$ should vanish. This in in contradiction with \cref{M2} and hence neither our interior nor any exact metric of which it could be an approximation can be a source of Kerr.

Now we focus on the Wahlquist family of metrics \cite{wahlquist1968isf}. That is to say to a stationary axisymmetric rigidly rotating perfect fluid with EOS $\epsilon+3p=\text{const.}$  and Petrov type D \cite{kramer1986rigidly,senovilla1987sap}.
In the non-static case, the conditions for Petrov type D for our interior family are given by \cref{deg2cond}. Then, in the case when $n=-2$ and \cref{deg2cond} is satisfied, $\bm{g}^-$ is of type D and our interior is an approximation to the Wahlquist family.

It must be noted nevertheless that, despite the fact that Wahlquist's family is a subcase of our general interior solution (as must be expected), the values  $m_2$ and $j_3$ imposed by the matching with the asymptotically flat exterior \textsl{do not} satisfy 
\cref{deg2cond}. Then, we recover and give an independent derivation of the known result that, within perturbation theory, Wahlquist's family can not correspond to an isolated source \cite{bradley2000wmc,sarnobat2006wes}.

A last comment. If one looks for stationary axisymmetric perfect fluid solutions with a static limit as candidates to represent the interior of a stellar model, consi\-de\-ring the Penrose chart of how a certain Petrov type can lead to another through degeneration, one should then only take into account those metrics whose Petrov type can lead to the types D and O corresponding to spherical symmetry \cite{senovilla1993contribucion}. A type N, rigidly rotating perfect fluid with barotropic EOS and $\epsilon+p\neq0$ can not be axisymmetric \cite{carminati1988type}, therefore we must discard types III and N, but all the rest should, to the best of our knowledge, be considered.
 It seems reasonable that a type II exact metric with the properties we demand can be approximated by our solutions. Nevertheless, the Petrov type II is not included among the possible types of our general interior metric and hence, we conjecture that \textsl{there is no stationary axisymmetric rigidly rotating perfect fluid metric with EOS $\epsilon+(1-n)p=\text{const.}$ of type II possessing a static limit and a surface of zero pressure.} This is in accordance with the weird fact that, even dropping the demand of zero pressure surface, 
 it has not been found any type II exact interior metric suitable to be part of a stellar model, while the harder field of type I solutions is not empty \cite{senovilla1993contribucion}.


\begin{acknowledgments}
The authors would like to thank J. Mart\'in and M. Mars for many helpful discussions. We are also grateful to the reviewers for several improvements to the original manuscript. This work was supported by grants FIS2009-07238 and FIS2012-30926. JEC thanks Junta de Castilla y Le\'on for grant EDU/1165/2007. The computation of the Petrov types was checked with the help of the tensor package \textsl{xAct}\cite{martin2008xperm} for \textsl{Mathematica} 
\end{acknowledgments}
 \appendix
\section{\texorpdfstring{$\mathcal{O}(\lambda^2,\Omega^3)$}{} metric components after Lichnerowciz matching}
Here we give both metrics written in the orthonormal cobasis associated to $\{t,r,\theta,\phi\}$. These result from the substitution of \cref{M0,M2,J1,J3,m0,m2,j1,lich1,lich2} in \cref{metExt,metInt}. The exterior components are

\begin{align}
 \gamma_{tt}&=-1+\lambda  \frac{1}{\eta}\left(2 -\frac{1}{\eta^2} \Omega ^2 P_2\right)+\lambda ^2 \frac{1}{\eta}\left\{\frac{28}{5}+\frac{2 n}{5} - \frac{2}{\eta}\right.\nonumber
\\
&\quad+\!\left. \Omega ^2 \left[\frac{16}{15}-\frac{4 n}{15} +\frac{1}{\eta^2}\left(-\frac{74}{35}+\frac{n}{7} + \frac{2}{\eta}\right) P_2\right]\right\},\\
\gamma_{t\phi}&=\lambda ^{3/2} \frac{1}{\eta^2}\left[\frac{4}{5}  \Omega  P^1_1+\Omega ^3 \left(\frac{2  }{3}P^1_1-\frac{2  }{7\eta^2}P^1_3\right)\right]\nonumber
\\
&\quad+\lambda ^{5/2}\frac{1}{\eta^2} \left\{\left(\frac{32}{7}+\frac{4 n}{35} -\frac{4 }{5\eta}\right) \Omega  P^1_1+\Omega ^3 \left[\left(\frac{352}{105}-\frac{6 n}{35} -\frac{2 }{3\eta}\right) P^1_1\right.\right.\nonumber
\\
&\quad+\!\left.\left.\frac{1}{\eta^2}\left(-\frac{992}{735}+\frac{22 n}{441} +\frac{12 }{35\eta}\right) P^1_3\right]\right\},
\\
\gamma_{rr}&=1+\lambda  \frac{1}{\eta}\left(2 -\frac{1}{\eta^2} \Omega ^2 P_2\right)+\lambda ^2 \frac{1}{\eta}\left\{\left(\frac{28}{5}+\frac{2 n}{5}\right) +2 \frac{1}{\eta}-\frac{16 }{35\eta^2}\right.\nonumber
\\
&\quad+\!\left.\Omega ^2 \left[\frac{16}{15}-\frac{4 n}{15}-\frac{8 }{105\eta^2}+\frac{1}{\eta^2}\left(-\frac{74}{35}+\frac{n}{7} -2 \frac{1}{\eta}+\frac{16 }{21\eta^2}\right) P_2\right]\right\},
\\
\gamma_{\theta\theta}&=1+\lambda  \frac{1}{\eta}\left(2 -\frac{1}{\eta^2} \Omega ^2 P_2\right)\nonumber
\\
&\quad+\lambda ^2 \frac{1}{\eta}\left\{\frac{28}{5}+\frac{2 n}{5} +\frac{1}{\eta}+\frac{8 }{35\eta^2}+\Omega ^2 \left[\frac{16}{15}-\frac{4 n}{15} +\frac{4 }{105\eta^2}\right.\right.\nonumber
\\
&\quad-\!\left.\left.\frac{1}{6\eta^3}+\frac{4 }{63\eta^4}+\frac{1}{\eta^2}\left(-\frac{74}{35}+\frac{n}{7} -\frac{5 }{6\eta}-\frac{4 }{9\eta^2}\right) P_2\right]\right\},
\\
\gamma_{r\theta}&=\lambda ^2 \Omega ^2\frac{1}{\eta^4}\left(\frac{1}{3}-\frac{16 }{63\eta}\right)  P^1_2
\\
\gamma_{\phi\phi}&=1+\lambda  \frac{1}{\eta}\left(2 -\frac{1}{\eta^2} \Omega ^2 P_2\right)\nonumber
\\
&\quad+\lambda ^2 \frac{1}{\eta}\left\{\left(\frac{28}{5}+\frac{2 n}{5}\right) +\frac{1}{\eta}+\frac{8}{35\eta^2}+\Omega ^2 \left[\frac{16}{15}-\frac{4 n}{15} +\frac{4 }{105\eta^2}\right.\right.\nonumber
\\
&\quad+\!\left.\left.\frac{1}{6\eta^3}-\frac{4 }{63\eta^4}+\frac{1}{\eta^2}\left(-\frac{74}{35}+\frac{n}{7} -\frac{7}{6\eta}-\frac{20 }{63\eta^2}\right) P_2\right]\right\}.
\end{align}
The results for the interior metric are
\begin{align}
 \gamma_{tt}&=-1+\lambda  \left(3-\eta ^2-\eta ^2 \Omega ^2 P_2\right)\nonumber
\\
&\quad+\lambda ^2 \left\{\frac{9}{2}+\frac{3 n}{4}-\left(1+\frac{n}{2}\right) \eta ^2+\left(\frac{1}{10}+\frac{3 n}{20}\right) \eta ^4+\Omega ^2 \left[1-\frac{n}{2}+\left({2}+{n}\right) \frac{\eta ^2}{3}\right.\right.\nonumber
\\
&\quad+\!\left.\left.\left(-\frac{3}{5}-\frac{n}{10}\right) \eta ^4+\eta ^2\left(-\frac{29}{35}-\frac{3 n}{14} +\left(\frac{5}{7}+\frac{5 n}{14}\right) \eta ^2\vphantom{\frac{{A^2}^2}{A2^2}}\right) P_2\right]\right\},
\\
\gamma_{t\phi}&=\lambda ^{3/2} \eta\left[\Omega\left(2  -\frac{6 \eta ^2}{5}\right)   P^1_1+\Omega ^3 \left(\frac{2  }{3}P^1_1-\frac{2 }{7}\eta ^2 P^1_3\right)\right]\nonumber
\\
&\quad+\lambda ^{5/2}\eta \left\{\left[\frac{49}{5}+\frac{n}{2} -\left(\frac{34}{5}+\frac{3 n}{5}\right) \eta ^2+\left(\frac{27}{35}+\frac{3 n}{14}\right) \eta ^4\right] \Omega  P^1_1\right.\nonumber
\\
&\quad+\!\left.\Omega ^3 \left[\left(\vphantom{\frac{{A^2}^2}{A2^2}}\frac{289}{105}-\frac{5 n}{14}  +\left(\frac{8}{15}+\frac{2 n}{5}\right) \eta ^2-\left(\frac{3}{5}+\frac{3 n}{14}\right) \eta ^4\right) P^1_1\right.\right.\nonumber
\\
&\quad+\!\left.\left.\eta ^2\left(-\frac{326}{245}-\frac{3 n}{49} +\left(\frac{34}{105}+\frac{n}{9}\right) \eta ^2\vphantom{\frac{{A^2}^2}{A2^2}}\right) P^1_3\right]\right\},
\\
\gamma_{rr}&=1+\lambda  \left(3-\eta ^2-\eta ^2 \Omega ^2 P_2\right)\nonumber
\\
&\quad+\lambda ^2 \left\{\frac{23}{2}+\frac{3 n}{4}+\left(-\frac{23}{5}-\frac{n}{2}\right) \eta ^2+\left(\frac{17}{70}+\frac{3 n}{20}\right) \eta ^4\right.\nonumber
\\
&\quad+\!\left.\Omega ^2 \left[\frac{5}{3}-\frac{n}{2}+\left(-\frac{14}{15}+\frac{n}{3}\right) \eta ^2+\left(\frac{9}{35}-\frac{n}{10}\right) \eta ^4\right.\right.\nonumber
\\
&\quad+\!\left.\left.\eta ^2\left(-\frac{17}{5}-\frac{3 n}{14} +\left(\frac{1}{21}+\frac{5 n}{14}\right) \eta ^2\vphantom{\frac{{A^2}^2}{A2^2}}\right) P_2\right]\right\},
\\
\gamma_{\theta\theta}&=1+\lambda  \left(3-\eta ^2-\eta ^2 \Omega ^2 P_2\right)\nonumber
\\
&\quad+\lambda ^2 \left\{\frac{23}{2}+\frac{3 n}{4}+\left(-\frac{26}{5}-\frac{n}{2}\right) \eta ^2+\left(\frac{37}{70}+\frac{3 n}{20}\right) \eta ^4\right.\nonumber
\\
&\quad+\!\left.\Omega ^2 \left[\frac{5}{3}-\frac{n}{2}+\left(-\frac{32}{35}+\frac{n}{3}\right) \eta ^2+\left(\frac{157}{630}-\frac{n}{10}\right) \eta ^4\right.\right.\nonumber
\\
&\quad+\!\left.\left.\eta ^2\left(-\frac{437}{105}-\frac{3 n}{14} +\left(\frac{97}{126}+\frac{5 n}{14}\right) \eta ^2\vphantom{\frac{{A^2}^2}{A2^2}}\right) P_2\right]\right\},
\\
\gamma_{r\theta}&=\lambda ^2 \Omega ^2\eta ^2\left(\frac{4 }{21}-\frac{\eta ^2}{9}\right)  P^1_2,
\\
\gamma_{\phi\phi}&=1+\lambda  \left(3-\eta ^2-\eta ^2 \Omega ^2 P_2\right)\nonumber
\\
&\quad+\lambda ^2 \left\{\frac{23}{2}+\frac{3 n}{4}-\left(\frac{26}{5}+\frac{n}{2}\right) \eta ^2+\left(\frac{37}{70}+\frac{3 n}{20}\right) \eta ^4\right.\nonumber
\\
&\quad+\!\left.\Omega ^2 \left[\frac{5}{3}-\frac{n}{2}+\left(-\frac{16}{105}+\frac{n}{3}\right) \eta ^2-\left(\frac{193}{630}+\frac{n}{10}\right) \eta ^4\right.\right.\nonumber
\\
&\quad+\!\left.\left.\eta ^2\left(-\frac{517}{105}-\frac{3 n}{14} +\left(\frac{167}{126}+\frac{5 n}{14}\right) \eta ^2\vphantom{\frac{{A^2}^2}{A2^2}}\right) P_2\right]\right\}.
\end{align}

\bibliography{bibliografMACROsepnames}

\end{document}